\documentclass[preprint,preprintnumbers,amsmath,amssymb,showpacs, doublespace]{revtex4}
\usepackage{graphicx}
\usepackage{dcolumn}
\usepackage{bm}
\usepackage{array,booktabs}

\def\be     {\begin{equation}}
\def\ee     {\end{equation}}
\def\bea        {\begin{eqnarray}}
\def\eea        {\end{eqnarray}}
\def\bnn    {\begin{eqnarray*}}
\def\enn    {\end{eqnarray*}}
\def\bc      {\begin{center}}
\def\ec      {\end{center}}
\def\bf      {\begin{figure}}
\def\ef      {\end{figure}}
\def\bpm     {\begin{pmatrix}}
\def\epm     {\end{pmatrix}}
\def\bvm    {\begin{vmatrix}}
\def\evm    {\end{vmatrix}}

\def\dag    {\dagger}
\def\f      {\frac}

\def\i      {\imath}
\def\bl      {\biggl}
\def\br     {\biggr}
\def\i         {\imath}
\def\limR    {\lim_{R \to 0}}
\def\Gam       {\Gamma}
\def\l          {\left}
\def\r          {\right}
\def\al         {\left |}
\def\ar         {\right |}
\def\gr        {\nabla}
\def\tr        {\textrm{tr}}
\def\h        {\hbar}

\def\d         {\mathcal{D}}

\def\tx        {\textrm}
\def\spin     {\textrm{spin}}
\def\intx      {\int d^{d}\bm{x}}
\def\intt       {\int^{\beta}_{0}d\tau}

\def\E          {\mathcal{E}}
\def\K          {\mathcal{K}}

\def\fs          {\fontsize}
\def\ns         {\normalsize}

\makeatletter
\renewcommand*\env@matrix[1][\arraystretch]{%
  \edef\arraystretch{#1}%
  \hskip -\arraycolsep
  \let\@ifnextchar\new@ifnextchar
  \array{*\c@MaxMatrixCols c}}
\makeatother

\begin{document}
\title{Interplay between disorder and inversion symmetry: Extreme enhancement of the mobility near the Weyl point in BiTeI}
\author{M. Sasaki$^{1,\ast}$, Kyoung-Min Kim$^{2,\ast}$, A. Ohnishi$^{1}$, M. Kitaura$^{1}$, N. Tomita$^{1}$, 
V. A. Kulbachinskii$^{3}$, Ki-Seok Kim$^{2}$, and Heon-Jung Kim$^{4,\dagger}$}
\affiliation{ $^{1}$Faculty of Science, Yamagata University, Kojirakawa, Yamagata 990-8560 Japan \\ $^{2}$ Department of Physics, Pohang University of Science and Technology (POSTECH), Pohang, Gyeongbuk 790-784, Republic of Korea \\ $^{3}$Physics Department, Moscow State University, Moscow 119899, Russia \\
$^{4}$Department of Physics, College of Natural Science, Daegu University, Gyeongbuk 712-714, Republic of Korea}
\date{\today}

\begin{abstract} 
We show experimental and theoretical evidence that BiTeI hosts a novel disordered metallic state named diffusive helical Fermi liquid (DHFL), characterized by a pair of concentric spin-chiral Fermi surfaces with negligible inter-valley scattering. Key experimental observations are extreme disparity of the mobility between inner and outer helical Fermi surfaces near the Weyl point and existence of the so called universal scaling behavior for the Hall resistivity. Although the extreme enhancement of the inner-Fermi-surface mobility near the Weyl point is quantitatively explained within the self-consistent Born approximation, the existence of universal scaling in the Hall resistivity shows its breakdown, implying necessity of mass renormalization in the inner Fermi-surface beyond the independent electron picture.
\end{abstract}

\maketitle 

\section{Introduction}
Change in the topological structure of the ground state, driven by disorders, has been intensively investigated recently and it is thought to be responsible for potential novel criticality which involves interplay of topological structures, disorders, and interactions \cite{Ruy12}. Such a change of topological structure is reflected in magneto-electrical transport phenomena such as anomalous Hall and spin Hall effects \cite{Nagaosa10}, negative longitudinal magnetoresistance \cite{Nielsen83,HJKim13}, chiral magnetic effect \cite{Fuku08}, and so on. Even if topology of the ground state is trivial, its anomalous geometric (local) structure described by the Berry curvature or determined by spin chirality can be also affected by disorders \cite{Xiao10}, showing an interesting variation of magneto-electrical transport, for example such as a crossover from weak anti-localization to weak localization driven by randomness. BiTeI may be an appropriate platform to investigate the interplay between Berry phase and disorder, originating from the unique electronic structure with broken inversion symmetry.

Inversion symmetry breaking in BiTeI splits a single degenerate band near the hexagonal face center of the Brillouin zone, referred to as the A point, into an inner band with a left-handed or ``positive'' spin-chiral configuration and an outer one with a right-handed or ``negative'', whose spin structures are intimately locked with momentum\cite{Ish11,Cre12,Lan12}. As a result, low energy physics of this inversion-symmetry-broken material is governed by two distinct Fermi surfaces when the Fermi energy $E_F$  lies near the band-touching point generated by the Rashba spin-orbit interaction. See Fig. 1(a). Dynamics of electrons on the inner Fermi surface (IFS) is described by the Weyl equation, exhibiting the change of the Fermi surface character from electron-like to hole-like across the Weyl point. Indeed, a nontrivial Berry phase of $\pi$ has been detected for both the IFS and outer Fermi surface (OFS) in the Shubnikov-de Haas measurements \cite{Mura13,Park}. Thus, this system is expected to show physics of a Weyl/Dirac metal \cite{Balents11} with interesting response to disorder \cite{Hosur12,Bjorn14,Roy14}. 

Up to now, however, this important point in BiTeI has been overlooked. Most electrical transport studies have focused on measurements at high magnetic fields to detect Shubnikov-de Haas or quantum oscillations \cite{Mura13,Park,Bell13,Martin13}. Probably, this is because Shubnikov-de Haas or quantum oscillations are considered to be  few experimental techniques to provide essential information about nontrivial Berry phase in this system \cite{Mura13,Park}. Another direction of research in BiTeI, in connection with its nontrivial topology is to induce a topological quantum phase transition and  a topological insulator under pressure, first proposed by Nagaosa and his colleagues \cite{Bah12}. Indeed, closing of the energy gap and some indirect signatures of the topological quantum phase transition were observed experimentally \cite{Xi13,Tran14}, but the nature of this topological critical point and of a topological insulator under pressure  are still elusive, particularly in the experimental point of view. 

In this paper, we investigate the interplay between the Berry phase and randomness in magneto-electrical transport properties of BiTeI. By analyzing Hall and magneto resistivity of Fermi-energy-tuned BiTeI single crystals, particularly at low magnetic fields, we reveal extreme disparity of the mobility values between IFS and OFS near the Weyl point, where the IFS mobility becomes colossally enhanced, intimately related with anti-localization in electrical transport. Based on the self-consistent Born approximation, we explain this disparity and ``divergent'' IFS mobility near the Weyl point quantitatively. We identify this fixed-point solution for BiTeI as a diffusive helical Fermi liquid, characterized by a pair of concentric spin-chiral Fermi surfaces with negligible inter-valley scattering. Our theoretical analysis indicates the existence of a crossover in the ``topological'' structure or geometric phase toward a conventional diffusive Fermi liquid when the stronger-disorder-enhanced inter-valley scattering destroys the spin-chiral property. However, we realize that this mean-field theory for disorders fails to describe the universal scaling in Hall resistivity, which is another main experiment result. We speculate that this failure in the self-consistent Born analysis implies the existence of mass renormalization of the IFS near the Weyl point, possibly resulting from enhanced interactions between electrons near the Weyl point.

Main experimental observations made on six Fermi-energy-tuned BiTeI single crystals are (1) an anomalous weak-field feature in Hall resistivity $\rho_H(B)$, (2) unconventional magnetic field $B$ dependence of magnetoresistance (MR), which is in stark contrast with the usual $B$-quadratic MR in a metal, and (3) a universal scaling of Hall resistivity.   The first experimental result is analyzed and understood based on a picture that two types of charge carriers exist in BiTeI: one with small mobility and the other with very large mobility. 
Indeed, we find that the overall negative slop in $\rho_H$($B$) is determined by electrons on the OFS
However, we also observe the deviation of $\rho_H$ from the linear dependence at the low $B$ region.
We assign it a contribution from the Weyl fermions in the IFS. 

The second result about MR is also consistent with the existence of two kinds of charge carriers in that the total electrical conductivity 
$\sigma_{total}$  in a magnetic field is decomposed by two channels of conduction given by $\sigma_{total} =\sigma_{OFS}+\sigma_{WF}$,
 where  $\sigma_{OFS}$ and  $\sigma_{WF}$ are the conductivity contributions of OFS and IFS, respectively. One can rewrite $\sigma_{total}$  into $\sigma_{total} = \sigma_c +\Delta\sigma^N_{out} +\Delta\sigma^N_{in} +\Delta\sigma_{WAL}$  with $\Delta\sigma_{WAL} \propto \sqrt{B}$, where $\sigma_c$,  $\Delta\sigma^N_{out(in)}$, and  $\Delta\sigma_{WAL}$  are the field-independent conductivity, the conductivity contribution of OFS (IFS), and the weak anti-localization correction in three dimensions (3D), respectively. The explicit form of each component will be given later. One important outcome in this MR analysis is the confirmation of the 3D weak anti-localization contribution in  $\sigma_{total}(B)$.
  The analysis of $\rho_H(B)$ and $\sigma_{total}$  enables us to extract separately the mobility values of the charge carriers in the OFS and IFS for all six samples. We plot the mobility values as a function of $E_F$ in Fig. 1(b). This data shows extreme disparity of the mobility values between IFS and OFS and ``divergent'' IFS mobility near the Weyl point. The detailed procedure how the IFS and OFS mobility values 
  are obtained will be presented in subsequent sections. 
  Here, we emphasize that mobility disparity and ``divergent'' IFS mobility near the Weyl point are determined not by the scattering time but by the transport time. As the transport time is a scattering time weighed more by backward scattering processes, chiral nature is more reflected in the transport time.

The rest of the paper is organized as follows. In Sec. II, we discuss the sample synthesis, magneto-electrical transport experiments, and analysis of the data in detail. In this section, we introduce two-carrier models for $\rho_H(B)$ and $\sigma_{total}(B)$, which is necessary to 
quantitatively explain the low-field features observed in  $\rho_H(B)$ and $\sigma_{total}(B)$. As an outcome of the analysis, we determine the OFS and IFS mobility values of six BiTeI single crystals with different $E_F$ [Fig. 1(b)].  In Sec. III, we calculate the IFS and OFS mobility values based on the Rashba model within the self-consistent Born approximation. Here we consider two different cases: one considers only the intra-valley forward scattering and the other includes both intra- and inter-valley scattering. It is revealed that the experimental IFS and OFS mobility values are quite well reproduced within this model in the absence of the inter-valley scattering or in the weak inter-valley scattering.  Besides, we predict how the ground state of BiTeI changes as the disorder increases by using renormalization group (RG) arguments. According to these, 
the inter-valley scattering smears out the spin-chirality with increasing disorder, leading to a topological crossover or a weak version of topological phase transition driven by disorder. This also accompanies the change of quantum correction in electrical transport from weak antilocalization to weak localization.   
Within this picture, the BiTeI single crystals which we investigate are in a weakly disordered region with negligible inter-valley scattering called diffusive helical Fermi liquid. In Sec. IV, we discuss implication of the experimental results based on the theory introduced in Sec. III. In fact, we find the existence of universal scaling in Hall resistivity from the experimental results. 
This scaling, however, is not reproduced within the self-consistent Born approximation. This necessitates mass renormalization in the IFS beyond the independent electron picture, especially near the Weyl point. We conclude in Sec. V with a brief discussion of our main results.       


\section{Experiment}
\subsection{Sample synthesis}
Single crystals of BiTeI were grown by a modified Bridgman method. We prepared more than 20 samples 
and tried to vary the carrier density $n$ by adding a small amount of extra Bi. 
As the amount of additionally inserted Bi is quite small, X-ray diffraction measurements do not detect any change of structure
 in the doped samples.
We selected six single crystals ($\#$1 $-$ $\#$6). 
All the as-grown single crystals were degenerate semiconductors, exhibiting a metallic behavior. 
Carrier densities $n$ were determined by the linear part of the Hall resistivity. 
Their signs are all negative, implying that dominant charge carriers are electrons, which presumably determined by the OFS. 
Carrier densities from the linear part were determined to be 0.10, 0.30, 0.35, 0.80, 3.9, and 
6.4$\times10^{20}$ cm$^{-3}$ for $\#$1 $-$ $\#$6, respectively. 
Estimated from the linear part of $\rho_H$, the Fermi energies from the bottom of the conduction band are 40, 90, 100, 170, 550, and 760 meV for all six samples. As the Weyl point is located at 113 meV from the bottom of the conduction band \cite{Bahramy11}, the former three (the latter three) should have positive (negative) charge carriers in the IFS. Later we will show that this, in fact is consistent with the sign change in the deviation of $\rho_H$ from the linear dependence.
Temperature dependence of the resistivity $\rho(T)$ are presented in Fig. 2, 
showing the overall decrease of the $\rho(T)$ curves with the increase of $n$. 
This behavior confirms that our samples are in the region of a typical degenerate semiconductor.
Specifying the distribution of the excess Bi and volatile I in the BiTeI samples can provide an important clue about the nature of disorder in this system,
 especially in connection with the results obtained in the present 
transport experiments. Even though we verified that our single crystals are homogeneous and uniform on macroscopic scale, probably because of small amount of excess Bi, it is considered that nanoscale inhomogeneity can still exist. What type of local disorders or defects can promote intra- or inter-valley scattering in BiTeI is a very important question which should be addressed in future studies. The six samples investigated in the present study are considered to be in weakly disordered region based on our final results, which suggests that effects of disorder are nearly equal at least in those samples for electrical transport.

\subsection{Analysis of magnetoresistivity and Hall resistivity}
The transverse MR = $(\rho(B) - \rho(0))/\rho(0)$ with $\rho(B)$ and $\rho(0)$, resistivity at $B$ and $B = 0$, respectively, 
and the Hall resistivity $\rho_H(B)$ for $\#1 - \#6$ are measured at 4.2 K and up to $B$ = 4 T. 
While the magnitude of MR is only few percent even at $B$ = 4 T for all samples, $\#1 - \#5$ show weak field anomalies, which deviate from the conventional $B$ quadratic behavior significantly, except for $\#6$ with the largest $n$, as presented in Fig. 3(a). In particular, the MR for $\#1$ 
possesses a pronounced dip in the weak field region. Even beyond the region of the dip, MR does not recover the $B$ quadratic behavior. 
The sample $\#1 - \#5$ exhibit essentially same features.

Hall resistivity curves are almost linear with negative slops, suggesting the existence of ``normal'' negative charge carriers. 
However, a more careful inspection for the low-field region reveals tiny weak-field anomalies, displayed in Fig. 3(b), 
which plots  the deviation $\Delta \rho_H$, where the overall linear dependence is subtracted from Hall resistvity $\rho_H(B)$. 
This data indicates that Hall resistivity curves deviate from the linearity significantly in the field region where a corresponding dip in MR is observed. While the deviations in $\#2 - \#5$ are confined 
for $-1$ T $<$ $B < 1$ T, they extend to $-4$ T $< B < 4$ T for $\#1$ and $\#6$. The shape of $\Delta \rho_H$ in Fig. 3(b)
 is reminiscent of the general case for the Hall resistivity with two types of charge carriers \cite{Wang14}. 
 In the limit that one mobility is much larger than the other, the formula at the low field region is simplified into
\be
\rho_H \approx \frac{1}{n_1ec}\frac{B}{1+(\mu_1B)^2}+\frac{B}{n_2ec},
\ee
where $n_1$ and $n_2$ are carrier densities with larger and smaller mobility, 
and $\mu_1$ is the larger mobility. If this simple expression explains the origin of the weak-field anomaly well, 
it suggests existence of a charge carrier with extremely high mobility, whose value corresponds to 
the maximum or minimum of $\Delta \rho_H$. Indeed, the first term of Eq. (1) fits the $\Delta \rho_H(B)$ data quite well, giving the mobility
value of the high mobility carrier as shown in Fig. 4.
We also observe that the sign of $\Delta \rho_H$ is positive for $\#1 - \#3$ 
and negative for $\#4 - \#6$, respectively, which implies that the charge carrier with extremely high mobility 
is hole-type for $\#1 - \#3$ and electron-type for  $\#4 - \#6$.

Considering the band structure of BiTeI near $E_F$, 
we assign the charge carrier with extremely high and the other one to be the Weyl fermion in the IFS and the OFS electron, respectively. 
While the OFS mobility  is estimated from the linear part of the Hall resistivity and the residual resistivity, the mobility of the Weyl fermions can be obtained from the fitting of $\Delta\rho_H$ to the first term of Eq. (1). Our analysis based on Eq. (1) turns out to explain $\Delta\rho_H$ in a quantitative level. The mobility values of the Weyl fermion and the OFS carrier are plotted for the six samples in Fig. 1(b) as a function of $E_F$.

It is appealing that the simple formula of Eq. (1) for the Hall resistivity explains the low-field region quite well. However, one might speculate that there must be anomalous Hall effect either intrinsic (Berry curvature) or extrinsic (side jump or skew scattering) \cite{Nagaosa10} because there are Weyl fermions in BiTeI. Although we cannot rule out the appearance of the extrinsic anomalous Hall effect, we strongly believe that the anomalous Hall effect induced by Berry curvature does not exist. The intrinsic anomalous Hall conductivity can be classified into two contributions, one of which results from the contribution of all states below the Fermi energy, given by the distance of momentum space between a pair of Weyl points \cite{Chen14,Goswami13}, and the other of which originates from the contribution of Fermi surfaces with Berry phase. The second is non-universal \cite{Haldane04,Xiao10,Nagaosa10}. Since a pair of Weyl points exists at the same momentum point, the first contribution vanishes. On the other hand, the second contribution from IFS and OFS may still exist, giving rise to an offset near the zero-field region. However, both contributions from the OFS and IFS will be cancelled because the sum of their Berry phases vanishes.

As in Hall resistivity, we also assume the existence of two conductivity channels. 
Then, the total contribution of electrical conductivity in BiTeI is given by $\sigma_{total}=\sigma_{OFS}+\sigma_{WF}$, 
where $\sigma_{OFS}=\frac{\sigma_{out}+\Delta\sigma^{out}_{WAL}}{\sigma_{out}^{-2}(\sigma_{out}+\Delta\sigma^{out}_{WAL})^2+\omega^2_{out}\tau^2_{out}}$
is the conductivity from the OFS and $\sigma_{WF}=\frac{\sigma_{in}+\Delta\sigma^{in}_{WAL}}{\sigma_{in}^{-2}(\sigma_{in}+\Delta\sigma^{in}_{WAL})^2+\omega^2_{in}\tau^2_{in}}$  is that from Weyl fermions of the IFS. These expressions can be derived from the Boltzmann-equation approach, where the role of the Berry phase is introduced into the Boltzmann equation via the weak anti-localization correction phenomenologically \cite{Ki14}.
$\sigma_{in(out)}$  and $\Delta\sigma_{WAL}^{in(out)}$   are the residual conductivity at zero magnetic field and the weak anti-localization correction, respectively.  $\omega_{in(out)}$ is the cyclotron frequency, and $\tau_{in(out)}$  is the transport time. 
Employing  $\Delta\sigma_{WAL}^{in(out)}=a_{in(out)}\sqrt{B}$  in three dimensions, we are allowed to assume $\sigma_{in(out)} >> \Delta\sigma_{WAL}^{in(out)}$  in the weak-field region. Then, these equations become simplified as follows, 
$\sigma_{OFS} \approx \frac{\sigma_{out}+\Delta\sigma^{out}_{WAL}}{1+\omega^2_{out}\tau^2_{out}} \approx \sigma_{out}^N+\Delta\sigma^{out}_{WAL}$  and
$\sigma_{WF} \approx \frac{\sigma_{in}+\Delta\sigma^{in}_{WAL}}{1+\omega^2_{in}\tau^2_{in}} \approx \sigma_{in}^N+\Delta\sigma^{in}_{WAL}$, respectively, where $\sigma_{out}^N = (\rho_{out}+A_{out}B^2)^{-1} \approx \rho_{out}^{-1}+\rho_{out}^{-2}A_{out}B^2$ 
with $\rho_{out} >> A_{out}B^2$  and $\sigma_{in}^N = (\rho_{in}+A_{in}B^2)^{-1}$ . 
The total magneto-electrical conductivity is finally written as $\sigma_{total} = \sigma_c + \Delta\sigma_{out}^N + \Delta\sigma_{in}^N + \Delta\sigma_{WAL}$   with $\Delta\sigma_{WAL} = \Delta\sigma_{WAL}^{out}+\Delta\sigma_{WAL}^{in}$,
 where all field-independent constants are expressed as $\sigma_c$.
 
The Fig. 5(a) show the decomposition of the magneto-electrical conductivity, $\Delta\sigma = \sigma_{total} - \sigma_c$  for the sample $\#1$. 
The sample $\#6$ with the highest $n$ is described only with $\Delta\sigma_{out}^{N}$  presumably because  $E_F$ is far away from the Weyl point. On the other hand, for other samples, all the other terms are necessary to describe the magneto-electrical conductivity properly. Performing successful decompositions, we isolate the weak anti-localization correction in Fig. 5(b), where all samples except for $\#6$ exhibit the scaling behavior with   $\sqrt{b}$ dependence, where $b$ is a dimensionless reduced magnetic field given by $b = \hbar \omega/E_F$, where 
$\omega$ is the cyclotron frequency. The existence of $\Delta\sigma_{WAL}$ for $\#1 - \#5$ 
justifies the validity of our data analysis.  

Our analysis on the magneto-electrical conductivity demonstrates existence of two types of charge carriers, one of which has an extremely high mobility, identified with Weyl fermions on the IFS, given by $\mu^2_{WF} = A_{in}/\rho_{in}$. Fig. 1(b) 
displays the mobility as a function of the Fermi energy  $E_F$, 
where the value of $\mu_{out} = \sqrt{A_{out}/\rho_{out}}$ [(black) open circles] is in the order of $0.01 \sim  0.03$ m$^2$/Vs 
while that of $\mu_{WF}$  [(red) open squares] is two or three orders of magnitude larger than $\mu_{out}$. In particular,  $\mu_{WF}$ touches its maximum when $E_F$  is closest to the Weyl point. The enhancement of $\mu_{WF}$ , compared to $\mu_{out}$, is partially a consequence of a reduced phase space available for the scattering in the IFS, which is an intrinsic property of the Weyl metal as derived in the following theoretical sections. It is noted that the mobility values deduced from $\Delta\rho_H$ are very similar to those from the MR analysis.

\section{Calculation of mobility values within the Rashba model}
\subsection{Effective model Hamiltonian} 

We start from the Rashba model with potential randomness:
\bnn
S[\bar{\Psi}_{i\alpha}(\tau,\bm{x}),\Psi_{i\alpha}(\tau,\bm{x})]&=&\f{1}{2}\intt\intx\br[\bar{\Psi}_{i\alpha}(\tau,\bm{x})\l(\partial_{\tau}-\f{\h^{2}\gr^{2}}{2m}-E_{F}\r)\Psi_{i\alpha}(\tau,\bm{x})\\
&&+\bar{\Psi}_{i\alpha}(\tau,\bm{x})\lambda_{R}\bm{\sigma}^{\spin}_{\alpha\beta}\cdot\l(\bm{E}\times(-\i\gr)\r)\Psi_{i\beta}(\tau,\bm{x})+\bar{\Psi}_{i\alpha}(\tau,\bm{x})V(\bm{x})\Psi_{i\alpha}(\tau,\bm{x})\bl] ,
\enn
where $\lambda_{R}$ is the Rashba coupling constant and $V(\bm{x})$ is a random potential. The indices of $``i"$ and $``\alpha"$ stand for time-reversal and spin component, repectively. ``Time-reversal symmetrized" basis is given by
\bnn
\Psi(\tau)=\bpm\psi(\tau)\\\i\sigma^{\spin}_{y}\psi^{*}(-\tau)\epm=\bpm\psi_{\uparrow}(\tau)\\\psi_{\downarrow}(\tau)\\\psi^{*}_{\downarrow}(-\tau)\\-\psi^{*}_{\uparrow}(-\tau)\epm ~ \& ~~ \bar{\Psi}(\tau)=\Psi^{\dag}(\tau)I^{\spin}\otimes\sigma^{\tr}_{z}=\bpm\psi^{*}_{\uparrow}(\tau)\\\psi^{*}_{\downarrow}(\tau)\\-\psi_{\downarrow}(-\tau)\\\psi_{\uparrow}(-\tau)\epm^{T}.
\enn

Taking into account the BiTeI band structure given by $\bm{E}=E\hat{z} ~ \l(\alpha_{R}=\lambda_{R}E\r)$ and moving on the momentum and frequency space, we obtain
\fs{10}{10}
\bnn
&&S[\bar{\Psi}_{i\alpha A(n)}(\bm{k}),\Psi_{i\alpha A(n)}(\bm{k})]\\
&=&\f{1}{2}\sum_{n}\int\f{d^{d}\bm{k}}{(2\pi)^{d}}\br[\bar{\Psi}_{i\alpha A(n)}(\bm{k})\l(-\i\omega_{n}+\f{\h^{2}\bm{k}^{2}}{2m}-E_{F}\r)\Psi_{i\alpha A(n)}(\bm{k})+\bar{\Psi}_{i\alpha n}(\bm{k})\alpha_{R}\l(k_{x}(\sigma_{y})_{\alpha\beta}-k_{y}(\sigma_{x})_{\alpha\beta}\r)\Psi_{j\beta A(n)}(\bm{k})\\
&&+\int\f{d^{d}\bm{q}}{(2\pi)^{d}}\bar{\Psi}_{i\alpha A(n)}(\bm{k}+\bm{q})V(\bm{q})\Psi_{i\alpha A(n)}(\bm{k})\bl] ,
\enn
\ns
where $A$ stands for ``retarded" ($\mathcal{R}$) or ``advanced" ($\mathcal{A}$). For example, $\mathcal{A}(n)$ corresponds to a negative frequency whose magnitude is $\al\omega_{n}\ar$. Diagonalizing this effective Rashba Hamiltonian based on the following momentum-dependent unitary matrix 
\fs{11}{11}
\bnn
U^{\dag}(\bm{k})I^{\spin}U(\bm{k})=I^{\spin} ~ \& ~ U(\bm{k})\l(k_{x}\sigma^{\spin}_{y}-k_{y}\sigma^{\spin}_{x}\r)U^{\dag}(\bm{k})=\sigma^{\spin}_{z}~\Rightarrow~U(\bm{k})=\f{1}{\sqrt{2}}\bpm e^{\i\f{\phi(\bm{k})}{2}}&-\i e^{-\i\f{\phi(\bm{k})}{2}}\\ e^{\i\f{\phi(\bm{k})}{2}}&\i e^{-\i\f{\phi(\bm{k})}{2}}\epm ,
\enn
\ns
we obtain 
\bnn
&&S[\bar{\Phi}_{i\alpha n}(\bm{k}),\Phi_{i\alpha n}(\bm{k})]\\
&=&\f{1}{2}\sum_{n}\int\f{d^{d}\bm{k}}{(2\pi)^{d}}\br[\bar{\Phi}_{i\alpha n}(\bm{k})\l(-\i\omega_{n}+\f{\h^{2}\bm{k}^{2}}{2m}-E_{F}\r)\Phi_{i\alpha n}(\bm{k})+\bar{\Phi}_{i\alpha n}(\bm{k})\alpha_{R}(\sigma^{\spin}_{z})_{\alpha\beta}\sqrt{k_{x}^{2}+k_{y}^{2}}\Phi_{j\beta n}(\bm{k})\\
&&~~~+\int\f{d^{d}\bm{q}}{(2\pi)^{d}}\bar{\Phi}_{i\alpha n}(\bm{k}+\bm{q})U_{\alpha\beta}(\bm{k}+\bm{q})V(\bm{q})U^{\dag}_{\beta\gamma}(\bm{k})\Phi_{i\gamma n}(\bm{k})\bl]
\enn
where $\Phi_{i\alpha n}(\bm{k})=U_{\alpha\beta}(\bm{k})\Psi_{i\beta n}(\bm{k})$ is an eigenfunction field and the index of $\alpha$ represents spin chirality, identified with ``$+$" or ``$-$". 

Performing the disorder average within the replica trick, the effective replicated Rashba action becomes
\fs{9.4}{9.4}
\bnn
&&S[\bar{\Phi}^{a}_{i\alpha n}(\bm{k}),\Phi_{i\alpha n}^{a}(\bm{k})]\\
&=&\f{1}{2}\sum_{n}\int\f{d^{d}\bm{k}}{(2\pi)^{d}}\br[\bar{\Phi}^{a}_{i\alpha n}(\bm{k})\l(-\i\omega_{n}+\f{\h^{2}\bm{k}^{2}}{2m}-E_{F}\r)\Phi^{a}_{i\alpha n}(\bm{k})+\bar{\Phi}^{a}_{i\alpha n}(\bm{k})\alpha_{R}\l(\sigma_{z}^{\spin}\r)_{\alpha\beta}\sqrt{k_{x}^{2}+k_{y}^{2}}\Phi^{a}_{i\beta n}(\bm{k})\bl]\\
&&-\sum_{nm}\int\f{d^{d}\bm{k}}{(2\pi)^{d}}\int\f{d^{d}\bm{k'}}{(2\pi)^{d}}\int\f{d^{d}\bm{q}}{(2\pi)^{d}}\f{\Gam}{8}\bar{\Phi}^{a}_{i\alpha n}(\bm{k}+\bm{q})M_{\alpha\alpha'}(\bm{k}+\bm{q},\bm{k})\Phi^{a}_{i\alpha' n}(\bm{k})\bar{\Phi}^{b}_{j\beta m}(\bm{k}'-\bm{q})M_{\beta\beta'}(\bm{k}'-\bm{q},\bm{k}')\Phi^{b}_{j\beta'm}(\bm{k}') ,
\enn
\ns
where the free energy is given by $\mathcal{F}=-T\limR\f{1}{R}\l(\int\d(\bar{\Phi}^{a}_{i\alpha},\Phi^{a}_{i\alpha})e^{-S[\bar{\Phi}^{a}_{i\alpha},\Phi^{a}_{i\alpha}]}-1\r)$. The product of unitary matrices is 
\fs{10}{10}
\bnn
M(\bm{k}+\bm{q},\bm{k})\equiv U(\bm{k}+\bm{q})U^{\dag}(\bm{k})= \bpm\cos{\l(\f{\phi(\bm{k}+\bm{q})-\phi(\bm{k})}{2}\r)}&\i\sin{\l(\f{\phi(\bm{k}+\bm{q})-\phi(\bm{k})}{2}\r)}\\\i\sin{\l(\f{\phi(\bm{k}+\bm{q})-\phi(\bm{k})}{2}\r)}&\cos{\l(\f{\phi(\bm{k}+\bm{q})-\phi(\bm{k})}{2}\r)}\epm.
\enn
\ns

\subsection{A self-consistent Born approximation}

We perform the Hubbard-Stratonovich transformation in the particle-hole singlet channel of $\Phi^{a}_{i\alpha n}\bar{\Phi}^{b}_{j\beta m}$, and obtain
\fs{10.5}{10.5}
\bnn
&&S[\bar{\Phi}^{a}_{i\alpha n}(\bm{k}),\Phi^{a}_{i\alpha n}(\bm{k});Q^{ab}_{ij;\alpha\beta;nm}(\bm{q})]\\
&=&\f{1}{2}\sum_{n}\int\f{d^{d}\bm{k}}{(2\pi)^{d}}\l[\bar{\Phi}^{a}_{i\alpha n}(\bm{k})\l(-\i\omega_{n}+\f{\h^{2}\bm{k}^{2}}{2m}-E_{F}\r)\Phi^{a}_{i\alpha n}(\bm{k})+\bar{\Phi}^{a}_{i\alpha n}(\bm{k})\alpha_{R}\l(\sigma^{\spin}_{z}\r)_{\alpha\beta}\sqrt{k_{x}^{2}+k_{y}^{2}}\Phi^{a}_{i\beta n}(\bm{k})\r]\\
&&+\sum_{nm}\int\f{d^{d}\bm{k}}{(2\pi)^{d}}\int\f{d^{d}\bm{q}}{(2\pi)^{d}}\br[-\f{\i}{2}\bar{\Phi}^{a}_{i\alpha n}(\bm{k}+\bm{q})M_{\alpha\alpha'}(\bm{k}+\bm{q},\bm{k})Q^{ab}_{ij;\alpha'\beta';nm}(\bm{q})M_{\beta'\beta}(\bm{k}-\bm{q},\bm{k})\Phi^{b}_{j\beta m}(\bm{k})\\
&&+\f{1}{2\Gam}M_{\alpha\alpha'}(\bm{k}+\bm{q},\bm{k})Q^{ab}_{ij;\alpha'\beta;nm}(\bm{q})M_{\beta\beta'}(\bm{k}-\bm{q},\bm{k})Q^{ba}_{ji;\beta'\alpha;mn}(-\bm{q})\bl].
\enn
\ns
Integrating over fermionic degrees of freedom, we obtain 
\fs{10.5}{10.5}
\bnn
&&S[Q^{ab}_{ij;\alpha\beta;A(n)B(m)}(\bm{q})]\\
&=&-\f{1}{2}\tr\tx{ln}\br[\delta^{ab}\delta_{AB}\l\{\delta_{ij}\delta_{\alpha\beta}\l(-\i\omega_{n}+\f{\h^{2}\bm{k}^{2}}{2m}-E_{F}\r)+\alpha_{R}(\sigma^{\spin}_{z})_{\alpha\beta}\sqrt{k_{x}^{2}+k_{y}^{2}}\r\}\\
&&-\i M_{\alpha\alpha'}(\bm{k}+\bm{q},\bm{k})Q^{ab}_{ij;\alpha'\beta';A(n)B(m)}(\bm{q})M_{\beta'\beta}(\bm{k}-\bm{q},\bm{k})\bl]\\
&&+\sum_{nm}\int\f{d^{d}\bm{k}}{(2\pi)^{d}}\int\f{d^{d}\bm{q}}{(2\pi)^{d}}\br[\f{1}{2\Gam}M_{\alpha\alpha'}(\bm{k}+\bm{q},\bm{k}) Q^{ab}_{ij;\alpha'\beta;A(n)B(m)}(\bm{q})M_{\beta\beta'}(\bm{k}-\bm{q},\bm{k})Q^{ba}_{ji;\beta'\alpha;B(m)A(n)}(-\bm{q})\bl].
\enn
\ns
Minimizing this effective free energy with respect to the matrix field $Q^{ab}_{ij;\alpha\beta;A(n)B(m)}(\bm{q})$, we obtain the saddle-point equation given by
%
%
\fs{10.5}{10.5}
\bnn
\f{2\i}{\Gam}Q^{ab}_{ij;\alpha\beta;A(n)B(m)}(\bm{q})=\tr\l[{G^{ab}_{ij;\alpha\beta;A(n)B(m)}}^{-1}(\bm{k})-\i M_{\alpha\alpha'}(\bm{k}+\bm{q},\bm{k})Q^{ab}_{ij;\alpha'\beta';A(n)B(m)}(\bm{q})M_{\beta'\beta}(\bm{k}-\bm{q},\bm{k})\r]^{-1} ,
\enn
\ns
where
\bnn
\l[G^{ab}_{ij;\alpha\beta;A(n)B(m)}(\bm{k})\r]^{-1}=\delta^{ab}\delta_{ij}\delta_{AB} \l\{\delta_{\alpha\beta}\l(-\i\omega_{n}+\f{\h^{2}\bm{k}^{2}}{2m}-E_{F}\r)+\alpha_{R}(\sigma^{\spin}_{z})_{\alpha\beta}\sqrt{k_{x}^{2}+k_{y}^{2}}\r\} 
\enn
is the inverse of the fermion Green's function.

\subsubsection{Saddle-point analysis I}

Focusing on the forward scattering described by $Q_{MF}=Q(\bm{0})$,
%
we obtain mean-field equations of
\bnn
\f{2\i}{\Gam}Q_{\scriptscriptstyle{++}}&=&\f{G_{\scriptscriptstyle{--}}^{-1}-\i Q_{\scriptscriptstyle{--}}}{\l(G_{\scriptscriptstyle{++}}^{-1}-\i Q_{\scriptscriptstyle{++}}\r)\l(G_{\scriptscriptstyle{--}}^{-1}-\i Q_{\scriptscriptstyle{--}}\r)+Q_{\scriptscriptstyle{+-}}Q_{\scriptscriptstyle{-+}}}\\
\f{2\i}{\Gam}Q_{\scriptscriptstyle{+-}}&=&\f{\i Q_{\scriptscriptstyle{+-}}}{\l(G_{\scriptscriptstyle{++}}^{-1}-\i Q_{\scriptscriptstyle{++}}\r)\l(G_{\scriptscriptstyle{--}}^{-1}-\i Q_{\scriptscriptstyle{--}}\r)+Q_{\scriptscriptstyle{+-}}Q_{\scriptscriptstyle{-+}}}\\
\f{2\i}{\Gam}Q_{\scriptscriptstyle{-+}}&=&\f{\i Q_{\scriptscriptstyle{-+}}}{\l(G_{\scriptscriptstyle{++}}^{-1}-\i Q_{\scriptscriptstyle{++}}\r)\l(G_{\scriptscriptstyle{--}}^{-1}-\i Q_{\scriptscriptstyle{--}}\r)+Q_{\scriptscriptstyle{+-}}Q_{\scriptscriptstyle{-+}}}\\
\f{2\i}{\Gam}Q_{\scriptscriptstyle{--}}&=&\f{G_{\scriptscriptstyle{++}}^{-1}-\i Q_{\scriptscriptstyle{++}}}{\l(G_{\scriptscriptstyle{++}}^{-1}-\i Q_{\scriptscriptstyle{++}}\r)\l(G_{\scriptscriptstyle{--}}^{-1}-\i Q_{\scriptscriptstyle{--}}\r)+Q_{\scriptscriptstyle{+-}}Q_{\scriptscriptstyle{-+}}},
\enn
where 
\bnn
G=\bpm G_{\scriptscriptstyle{++}}&0\\0&G_{\scriptscriptstyle{--}}\epm ~ \& ~ Q=\bpm Q_{\scriptscriptstyle{++}}&Q_{\scriptscriptstyle{+-}}\\Q_{\scriptscriptstyle{-+}}&Q_{\scriptscriptstyle{--}}\epm  
\enn
and the forward scattering doesn't change spin orientations, resulting in $M(\bm{k},\bm{k})=I$. It is straightforward to see $Q_{\scriptscriptstyle{+-}}=0$ due to spin chirality. 
%
%
Then, we reach the following expression
%
%
\fs{10.5}{10.5}
\bnn
\f{2\i}{\Gam}Q^{ab}_{ij;\pm\pm;A(n)A(n)}(\bm{0})=\int\f{d^{d}\bm{k}}{(2\pi)^{d}}\f{1}{\delta^{ab}\delta_{AB}\l\{\delta_{ij}\l(-\i\omega_{n}+\f{\h^{2}\bm{k}^{2}}{2m}-E_{F}\r)\pm\alpha_{R}\sqrt{k_{x}^{2}+k_{y}^{2}}\r\}-\i Q^{ab}_{ij;\pm\pm;A(n)A(n)}(\bm{0})}.
\enn
\ns

In order to solve the above equation, we introduce the mean-field ansatz of
\bnn
(Q_{\tx{MF}})^{ab}_{ij;\alpha\beta;AB}=\f{\pi}{2}N_{F}\Gam\delta^{ab}\delta_{ij}\delta_{\alpha\beta}F_{\alpha}(r)\Lambda_{AB} ,
\enn
where $N_{F}=mk_{F}/2\pi^{2}\h^{2}$ is the density of the states (without the factor $2$ of spin-degeneracy) at the Fermi level with $\alpha_{R}=0$ and $\Lambda_{AB}=\tx{diag}(1,-1)$ is the diagonal matrix for the retarded and advanced sectors. $F_{\alpha}(r)$ is a function of $r=\f{2\h^{2}E_{F}}{m\alpha_{R}^{2}}$, regarded as an order parameter to be determined from this self-consistent equation. Considering $\alpha=\beta=+$ and $A=B=\mathcal{R}$, we obtain
\bnn
\i \pi N_{F}F_{+} = \int\f{d^{3}\bm{k}}{(2\pi)^{3}}\f{1}{-\i\omega_{n}+\f{\h^{2}\bm{k}^{2}}{2m}-E_{F}+\alpha_{R}\sqrt{k_{x}^{2}+k_{y}^{2}}-\f{\i\pi N_{F}\Gam F_{+}}{2}}.
\enn
Since this integrand is not rotationally invariant along the $z-$direction, we need to be cautious for the $k_{z}$ integration. We will not show this procedure. Performing the momentum integration,  we find
\bnn
&& F_{+}(r)
= \f{\pi}{2}\f{1}{\sqrt{r}\sqrt{1+r}}\br[\f{8}{3}+2r-\f{4}{3}\sqrt{r+1}\br\{\l(r+2\r)\E\l(\f{1}{r+1}\r)-r\K\l(\f{1}{r+1}\r)\bl\}\bl],
\enn
where $\K(x)$ and $\E(x)$ are complete elliptic integrals of the first kind and the second kind \cite{Mathematica}. In the same way, we find 
\bnn
&& F_{-}(r)= \f{\pi}{2}\f{1}{\sqrt{r}\sqrt{1+r}}\br[\f{8}{3}+2r+\f{4}{3}\sqrt{r+1}\br\{\l(r+2\r)\E\l(\f{1}{r+1}\r)-r\K\l(\f{1}{r+1}\r)\bl\}\bl],
\enn
where the sign in front of the elliptic integrals has been changed. As a result, two kinds of scattering times are given by
%
%
\bnn
\tau_{+}=\f{1}{2Q_{+}(r)}=\f{1}{\pi N_{F}\Gam F_{+}(r)} ~ \& ~ \tau_{-}=\f{1}{2Q_{-}(r)}=\f{1}{\pi N_{F}\Gam F_{-}(r)}
\enn
for inner and outer Fermi surfaces, respectively. Considering that the scattering time is expressed as $\tau=\f{1}{\pi N_{F}\Gam}$ for the normal diffusive Fermi-liquid state, one may regard that the appearance of the additional factor of $F_{\pm}(r)$ results from the presence of the Rashba spin-orbit coupling, modifying the density of states for inner and our Fermi surfaces, respectively. Finally, we obtain diffusion constants, given by
\bnn
D_{\pm}=\h v_{F}^{2}\tau_{\pm} = \f{2\pi\alpha_{R}\h^{3}}{m^{2}\Gam}\f{(1+r)}{\sqrt{r}F_{\pm}(r)}.
\enn

Although we did not show the integration procedure in a detail, these diffusion coefficients are justified only when $r \geq 0$. Since there are no density of states for the inner Fermi surface (FIG. \ref{FS_3d}), we divide this case when the Fermi energy is below the Weyl point from the other. As a result, we find the general formula valid for both cases of $r \geq 0$ and $r < 0$, given by
%
%
\fs{8.5}{8.5}
\bnn
D_{\pm}(r)=\f{4\alpha_{R}\h^{3}}{m^{2}\Gam}\f{(1+r)^{\f{3}{2}}}{\tx{Re}\l[\f{8}{3}+2r-\f{4}{3}\sqrt{r+1}\l\{(r+2)\E\l(\f{1}{r+1}\r)-r\K\l(\f{1}{r+1}\r)\r\}\r]-\Theta(-r)\l(\f{8}{3}-\f{8}{3}\sqrt{1-\al r\ar}+\f{2}{3}\al r\ar\sqrt{1-\al r\ar}-2\al r\ar\r)} .
\enn
\ns
In order to compare our analytic expressions with the experimental data, we need to obtain the mobility. Resorting to the Einstein's relation $D=\mu k_{B}T/e$, we have
\fs{4.7}{2}
\bnn
\mu_{\pm}(E_{F})=A\f{(1+bE_{F})^{\f{3}{2}}}{\tx{Re}\l[\f{8}{3}+2bE_{F}-\f{4}{3}\sqrt{bE_{F}+1}\l\{(bE_{F}+2)\E\l(\f{1}{bE_{F}+1}\r)-bE_{F}\K\l(\f{1}{bE_{F}+1}\r)\r\}\r]-\Theta(-bE_{F})\l(\f{8}{3}-\f{8}{3}\sqrt{1-\al bE_{F}\ar}+\f{2}{3}\al bE_{F}\ar\sqrt{1-\al bE_{F}\ar}-2\al bE_{F}\ar\r)} ,
\enn
\ns
where $A=\f{2\alpha_{R}\h^{3}}{m^{2}\Gam ek_{B}T}$ and $b=\f{2\h^{2}}{m\alpha_{R}^{2}}$. In the experiment, the Weyl point was observed at 113 meV from the bottom of the conduction band mimnimum. In our model, we set $E_{W}=0$ and the conduction band minimum is $-\f{m\alpha_{R}^{2}}{2\h^{2}}$, so $b=\f{2\h^{2}}{m\alpha_{R}^{2}}=\f{1}{0.113eV}\simeq$ 8.85 $(eV)^{-1}$. Based on the formula with this value, we fit the experimental data and obtain the result of FIG. \ref{diffusion_constant_3d}, where $A=$ 0.984 $[m^{2}/Vs]$. 

%
%

\subsubsection{Saddle-point analysis II}

Previously, we did not take into account effects of inter-valley scattering. Taking $\bm{q}=-2\bm{k}-\bm{a}$ where $\bm{a}=\f{2m\alpha_{R}}{\h^{2}}\f{k_{x}\hat{x}+k_{y}\hat{y}}{\sqrt{k_{x}^{2}+k_{y}^{2}}}$, the effective Rashba action becomes
\fs{10}{10}
\bnn
&&S[Q^{ab}_{ij;\alpha\beta;A(n)B(m)}(-2\bm{k}-\bm{a})]\\
&=&-\f{1}{2}\tr\tx{ln}\br[\delta^{ab}\delta_{AB}\l\{\delta_{ij}\delta_{\alpha\beta}\l(-\i\omega_{n}+\f{\h^{2}\bm{k}^{2}}{2m}-E_{F}\r)+\alpha_{R}(\sigma^{\spin}_{z})_{\alpha\beta}\sqrt{k_{x}^{2}+k_{y}^{2}}\r\}\\
&&-\i M_{\alpha\alpha'}(-\bm{k}-\bm{a},-\bm{k})Q^{ab}_{ij;\alpha'\beta';A(n)B(m)}(-2\bm{k}-\bm{a})M_{\beta'\beta}(\bm{k}+\bm{a},\bm{k})\bl]\\
&&+\sum_{nm}\int\f{d^{d}\bm{k}}{(2\pi)^{d}}\f{1}{2\Gam}\br[M_{\alpha\alpha'}(-\bm{k}-\bm{a},-\bm{k})Q^{ab}_{ij;\alpha'\beta;A(n)B(m)}(-2\bm{k}-\bm{a})M_{\beta\beta'}(\bm{k}+\bm{a},\bm{k}) Q^{ba}_{ji;\beta'\alpha;B(m)A(n)}(2\bm{k}+\bm{a})\bl].
\enn
\ns
Since $\bm{k}+\bm{a}$ and $\bm{k}$ are in the same direction on the $xy-$plane, we still have $M(-\bm{k}-\bm{a},-\bm{k})=M(\bm{k}+\bm{a},\bm{k})=I$. Then, the above expression is simplified as follows
\fs{10.5}{10.5}
\bnn
&&S[Q^{ab}_{ij;\alpha\beta;A(n)B(m)}(-2\bm{k}-\bm{a})]\\
&=&-\f{1}{2}\tr\tx{ln}\br[\delta^{ab}\delta_{AB}\l\{\delta_{ij}\delta_{\alpha\beta}\l(-\i\omega_{n}+\f{\h^{2}\bm{k}^{2}}{2m}-E_{F}\r)+\alpha_{R}(\sigma^{\spin}_{z})_{\alpha\beta}\sqrt{k_{x}^{2}+k_{y}^{2}}\r\}-\i Q^{ab}_{ij;\alpha\beta;nm}(-2\bm{k}-\bm{a})\bl]\\
&&+\sum_{nm}\f{1}{2\Gam}\br[Q^{ab}_{ij;\alpha\beta;A(n)B(m)}(-2\bm{k}-\bm{a})Q^{ba}_{ji;\beta\alpha;B(m)A(n)}(2\bm{k}+\bm{a})\bl].
\enn
\ns
Unfortunately, this effective action is not diagonal in the presence of such a $Q(2\bm{k}+\bm{a})$ matrix. We can resolve this difficulty, choosing a better basis as
\fs{9.5}{9.5}
\bnn
&&\bar{\phi}(\bm{k})\l[G^{-1}(\bm{k})\r]\phi(\bm{k})+\bar{\phi}(-\bm{k}-\bm{a})\l[G^{-1}(-\bm{k}-\bm{a})\r]\phi(-\bm{k}-\bm{a})+\bar{\phi}(-\bm{k}-\bm{a})\l[-\i Q(-2\bm{k}-\bm{a})\r]\phi(\bm{k})\\
&&+\bar{\phi}(\bm{k})\l[-\i Q(2\bm{k}+\bm{a})\r]\phi(-\bm{k}-\bm{a})\\
&=&\bpm\bar{\phi}(\bm{k}),&\bar{\phi}(-\bm{k}-\bm{a})\epm\bpm G^{-1}(\bm{k})&-\i Q(\bm{2\bm{k}+\bm{a}})\\-\i Q(-2\bm{k}-\bm{a})&G^{-1}(-\bm{k}-\bm{a})\epm\bpm\phi(\bm{k})\\\phi(-\bm{k}-\bm{a})\epm\\
&=&\bpm\bar{\phi}_{+}(\bm{k})\\\bar{\phi}_{-}(\bm{k})\\\bar{\phi}_{+}(-\bm{k}-\bm{a})\\\bar{\phi}_{-}(-\bm{k}-\bm{a})\epm^{T}\bpm G^{-1}_{\scriptscriptstyle{++}}(\bm{k})&0&0&-\i Q_{\scriptscriptstyle{+-}}(\bm{2\bm{k}+\bm{a}})\\0&G^{-1}_{\scriptscriptstyle{--}}(\bm{k})&-\i Q_{\scriptscriptstyle{-+}}(2\bm{k}+\bm{a})&0\\0&-\i Q_{\scriptscriptstyle{+-}}(-2\bm{k}-\bm{a})&G^{-1}_{\scriptscriptstyle{++}}(-\bm{k}-\bm{a})&0\\-\i Q_{\scriptscriptstyle{-+}}(-2\bm{k}-\bm{a})&0&0&G^{-1}_{\scriptscriptstyle{--}}(-\bm{k}-\bm{a})\epm\bpm\phi_{+}(\bm{k})\\\phi_{-}(\bm{k})\\\phi_{+}(-\bm{k}-\bm{a})\\\phi_{-}(-\bm{k}-\bm{a})\epm.
\enn
\ns

This expanded matrix can be made to be a block-diagonal form, so we are allowed to consider two components only:
\bnn
\bpm\bar{\phi}_{+}(\bm{k}),&\bar{\phi}_{-}(-\bm{k}-\bm{a})\epm\bpm G^{-1}_{\scriptscriptstyle{++}}(\bm{k})&-\i Q_{\scriptscriptstyle{+-}}(\bm{2\bm{k}+\bm{a}})\\-\i Q_{\scriptscriptstyle{-+}}(-2\bm{k}-\bm{a})&G^{-1}_{\scriptscriptstyle{--}}(-\bm{k}-\bm{a})\epm\bpm\phi_{+}(\bm{k})\\\phi_{-}(-\bm{k}-\bm{a})\epm.
\enn
As a result, we find self-consistent equations for inter-valley scattering
\fs{11}{11}
\bnn
\f{2\i}{\Gam}\cdot0&=&\int\f{d^{3}\bm{k}}{(2\pi)^{3}}\f{G_{\scriptscriptstyle{--}}^{-1}(-\bm{k}-\bm{a})}{G^{-1}_{\scriptscriptstyle{++}}(\bm{k})G^{-1}_{\scriptscriptstyle{--}}(-\bm{k}-\bm{a})+Q_{\scriptscriptstyle{+-}}(2\bm{k}+\bm{a})Q_{\scriptscriptstyle{-+}}(-2\bm{k}-\bm{a})}\\
\f{2\i}{\Gam}Q_{\scriptscriptstyle{+-}}(2\bm{k}+\bm{a})&=&\int\f{d^{3}\bm{k}}{(2\pi)^{3}}\f{\i Q_{\scriptscriptstyle{+-}}(2\bm{k}+\bm{a})}{G_{\scriptscriptstyle{++}}^{-1}(\bm{k})G_{\scriptscriptstyle{--}}^{-1}(-\bm{k}-\bm{a})+Q_{\scriptscriptstyle{+-}}(2\bm{k}+\bm{a})Q_{\scriptscriptstyle{-+}}(-2\bm{k}-\bm{a})}\\
\f{2\i}{\Gam}Q_{\scriptscriptstyle{-+}}(-2\bm{k}-\bm{a})&=&\int\f{d^{3}\bm{k}}{(2\pi)^{3}}\f{\i Q_{\scriptscriptstyle{-+}}(-2\bm{k}-\bm{a})}{G_{\scriptscriptstyle{++}}^{-1}(\bm{k})G_{\scriptscriptstyle{--}}^{-1}(-\bm{k}-\bm{a})+Q_{\scriptscriptstyle{+-}}(2\bm{k}+\bm{a})Q_{\scriptscriptstyle{-+}}(-2\bm{k}-\bm{a})}\\
\f{2\i}{\Gam}\cdot0&=&\int\f{d^{3}\bm{k}}{(2\pi)^{3}}\f{G_{\scriptscriptstyle{++}}^{-1}(\bm{k})}{G_{\scriptscriptstyle{++}}^{-1}(\bm{k}) G_{\scriptscriptstyle{--}}^{-1}(-\bm{k}-\bm{a})+Q_{\scriptscriptstyle{+-}}(2\bm{k}+\bm{a})Q_{\scriptscriptstyle{-+}}(-2\bm{k}-\bm{a})},
\enn
\ns
where $Q_{\scriptscriptstyle{++}(\scriptscriptstyle{--})}(\pm(2\bm{k}+\bm{a}))$ turn out to vanish due to spin chirality. Note that $G_{\scriptscriptstyle{++}}(\bm{k})$ being on the Fermi surface means $G_{\scriptscriptstyle{--}}(-\bm{k}-\bm{a})=G_{\scriptscriptstyle{--}}(\bm{k}+\bm{a})$ should also be on the Fermi surface.
%
%
Thus, $Q_{\scriptscriptstyle{+-}}$ doesn't have to vanish in this case. 
%
%
Linearizing the energy spectrum around the inner Fermi surface, we obtain
\bnn
\f{2\i}{\Gam}Q_{\scriptscriptstyle{+-}}&=&\int^{2\pi}_{0}\f{d\phi}{2\pi}\int^{\pi}_{0}\f{d\theta J_{+}(\sin{\theta})}{2\pi}\int^{\infty}_{-\infty}\f{d\varepsilon}{2\pi}\f{\i Q_{\scriptscriptstyle{+-}}}{J(\sin{\theta})\l(\h v_{F}\varepsilon\r)^{2}+\al Q_{\scriptscriptstyle{+-}}\ar^{2}}
\enn
where $J_{+}(\sin{\theta})$ is a Jacobian factor from expanding $\bm{k}$ around the inner Fermi surface and $J(\sin{\theta})$ is a Jacobian factor from connecting the integral on the outer Fermi surface to the integral on the inner fermi surface. Other two equations are satisfied automatically, identically zero. Straightforward calculations give rise to the final expression
%
%
\bnn
\al Q_{\scriptscriptstyle{+-}}\ar=\f{\Gam}{2}\f{1}{2\h v_{F}}\l(\f{m\alpha_{R}}{2\pi\h^{4}}\r)^{2}\l(ar+br^{2}+cr^{3}\r)=\f{\pi N_{F}\Gam}{4}\sqrt{\f{r}{1+r}}\l(a+br+cr^{2}\r)
\enn
where $a=9.42\times10^{-3},~b=2.36\times10^{-1}$ and $c=-3.62\times10^{-2}$.

In the presence of the off-diagonal term of $Q_{\scriptscriptstyle{+-}}$, the fermion propagator is altered as follows
\fs{9}{9}
\bnn
\bpm G^{-1}_{\scriptscriptstyle{++}}-\i Q_{\scriptscriptstyle{++}}&-\i Q_{\scriptscriptstyle{+-}}\\-\i Q_{\scriptscriptstyle{-+}}&J\l(G_{\scriptscriptstyle{--}}^{-1}-\i Q_{\scriptscriptstyle{--}}\r)\epm^{-1}=\f{1}{\l(G_{\scriptscriptstyle{++}}^{-1}-\i Q_{\scriptscriptstyle{++}}\r)J\l(G^{-1}_{\scriptscriptstyle{--}}-\i Q_{\scriptscriptstyle{--}}\r)+\al Q_{\scriptscriptstyle{+-}}\ar^{2}}\bpm J\l(G_{\scriptscriptstyle{--}}^{-1}-\i Q_{\scriptscriptstyle{--}}\r)&\i Q_{\scriptscriptstyle{+-}}\\\i Q_{\scriptscriptstyle{-+}}&G_{\scriptscriptstyle{++}}^{-1}-\i Q_{\scriptscriptstyle{++}}\epm.
\enn
\ns
Then, the effective inner-fermion propagator is given by
\bnn
G_{\scriptscriptstyle{++}\tx{eff}}^{-1}(\bm{k})&=&\f{1}{G^{-1}_{\scriptscriptstyle{++}}(\bm{k})-\i Q_{\scriptscriptstyle{++}}(\bm{0})+\f{\al Q_{\scriptscriptstyle{+-}}(2\bm{k}+\bm{a})\ar^{2}}{J(\Omega)\l(G_{\scriptscriptstyle{--}}^{-1}(-\bm{k}-\bm{a})-\i Q_{\scriptscriptstyle{--}}(\bm{0})\r)}}.
\enn
Taking $\bm{k}=\bm{k}^{+}_{F}$, where $\bm{k}_{F}^{+}$ is the Fermi momentum of the inner Fermi surface,
%
%
we find 
\bnn
Q_{\scriptscriptstyle{++}\tx{eff}}(\bm{0})=Q_{\scriptscriptstyle{++}}(\bm{0})\l[1-\f{\al Q_{\scriptscriptstyle{+-}}(2\bm{k}_{F}^{+}+\bm{a})\ar^{2}}{Q_{\scriptscriptstyle{++}}(\bm{0})Q_{\scriptscriptstyle{--}}(\bm{0})}\r]=\f{\pi}{2}N_{F}\Gam F_{+}\l[1-\f{r}{4(1+r)}\f{\l(a+br+cr^2\r)^{2}}{F_{+}F_{-}}\r].
\enn
Accordingly, scattering times are modified as
\bnn
\tau_{\pm\tx{eff}}=\f{1}{2Q_{\scriptscriptstyle{\pm\pm}\tx{eff}}}=\f{1}{2\f{\pi}{2}N_{F}\Gam F_{\pm}\l[1-\f{r}{4(1+r)}\f{\l(a+br+cr^2\r)^{2}}{F_{+}F_{-}}\r]}.
\enn
As a result, diffusion constants are given by
\bnn
D_{\pm\tx{eff}}(r)=\h v_{F}^{2}\tau_{\pm\tx{eff}} = \f{2\pi\alpha_{R}\h^{3}}{m^{2}\Gam}\f{1+r}{\sqrt{r}}\f{1}{F_{\pm}\l(1-\f{r}{4(1+r)}\f{\l(a+br+cr^2\r)^{2}}{F_{+}F_{-}}\r)}
\enn
where $F_{\pm}(r)$ are 
\bnn
F_{\pm}(r)&=&\f{\pi}{2}\f{1}{\sqrt{r}\sqrt{1+r}}\tx{Re}\l[\f{8}{3}+2r-\f{4}{3}\sqrt{r+1}\l\{(r+2)\E\l(\f{1}{r+1}\r)-r\K\l(\f{1}{r+1}\r)\r\}\r]\\
&&-\Theta(-r)\l(\f{8}{3}-\f{8}{3}\sqrt{1-\al r\ar}+\f{2}{3}\al r\ar\sqrt{1-\al r\ar}-2\al r\ar\r) ,
\enn
the same as before. These results are summarized and compared to experiments in FIG. \ref{diffusion_constant_3d_with_offQ}, where this fixed-point solution (the presence of the off-diagonal component) turns out to reduce the mobilities of both fermions at inner and outer Fermi surfaces.

\subsection{Two different cases within the self-consistent Born approximation}

Previously, we considered two types of solutions for the fermion Green's function within the Born approximation: One contains effects of only the intra-valley forward scattering and the other introduces both effects of intra- and inter-valley scattering into the fermion Green's function. It is natural to expect that the former solution would be justified when effects of disorder scattering are not strong. On the other hand, the second solution is expected to be realized when disorder scattering becomes more relevant than the first case. A crucially different point between these two solutions lies in spin chirality. Intra-valley scattering preserves the spin chirality while inter-valley scattering destroys it. As a result, we predict that the weak antilocalization turns into the weak localization in the presence of inter-valley scattering, increasing disorder strength. This crossover behavior may be regarded as a weak version of a topological phase transition driven by disorder although BiTeI is a topologically trivial metallic state.

\subsection{Self-consistent Born approximation as a fixed-point solution}

The solution based on the self-consistent Born approximation can be regarded as an effective theory for the corresponding diffusive Fermi-liquid fixed point. In order to understand this statement, we consider the following renormalization group equation for disorder strength up to one-loop order
%
%
\bnn
\f{d\Gam_{ss'}}{dt}=\Gam_{ss'}-\Gam_{ss_{1}}\mathcal{C}_{s_{1}s_{2}}\Gam_{s_{2}s'} .
\enn
$\Gam_{\scriptscriptstyle{++}(\scriptscriptstyle{--})}$ is the scattering rate or variance within the inner (outer) Fermi surface and $\Gam_{\scriptscriptstyle{+-}(\scriptscriptstyle{-+})}$ is that between the inner and outer Fermi surfaces. The first term ensures the relevance of disorder scattering in the tree level when there is a Fermi surface. Such relevant disorder scattering becomes weak through quantum fluctuations, where the disorder potential is screened by particle-hole excitations. $\mathcal{C}_{s_{1}s_{2}}$ are positive constants, computed in quantum corrections of the one-loop level. $t$ is the renormalization-group transformation scale.

These renormalization group equations can be rewritten as follows
%
%
\bnn
\f{d\Gam_{\scriptscriptstyle{++}}}{dt}&=&\Gam_{\scriptscriptstyle{++}}-\Gam_{\scriptscriptstyle{++}}C_{\scriptscriptstyle{++}}\Gam_{\scriptscriptstyle{++}} -\Gam_{\scriptscriptstyle{+-}}C_{\scriptscriptstyle{--}} \Gam_{\scriptscriptstyle{-+}}\\
\f{d\Gam_{\scriptscriptstyle{+-}}}{dt}&=&\Gam_{\scriptscriptstyle{+-}}-\Gam_{\scriptscriptstyle{++}}C_{\scriptscriptstyle{++}}\Gam_{\scriptscriptstyle{+-}} -\Gam_{\scriptscriptstyle{+-}}C_{\scriptscriptstyle{--}} \Gam_{\scriptscriptstyle{--}}\\
\f{d\Gam_{\scriptscriptstyle{-+}}}{dt}&=&\Gam_{\scriptscriptstyle{-+}}-\Gam_{\scriptscriptstyle{-+}}C_{\scriptscriptstyle{++}}\Gam_{\scriptscriptstyle{++}} -\Gam_{\scriptscriptstyle{--}}C_{\scriptscriptstyle{--}} \Gam_{\scriptscriptstyle{-+}}\\
\f{d\Gam_{\scriptscriptstyle{--}}}{dt}&=&\Gam_{\scriptscriptstyle{--}}-\Gam_{\scriptscriptstyle{--}}C_{\scriptscriptstyle{--}}\Gam_{\scriptscriptstyle{--}} -\Gam_{\scriptscriptstyle{-+}}C_{\scriptscriptstyle{++}} \Gam_{\scriptscriptstyle{+-}} .
\enn  
Fixed points are determined by $d \Gamma_{ss'} / d t = 0$, resulting in
\bnn
&&\Gam_{\scriptscriptstyle{++}}-\Gam_{\scriptscriptstyle{++}}^{2}C_{\scriptscriptstyle{++}}-\Gam_{\scriptscriptstyle{+-}}\Gam_{\scriptscriptstyle{-+}}C_{\scriptscriptstyle{--}} =0\\
&&\Gam_{\scriptscriptstyle{+-}}\l(1-\Gam_{\scriptscriptstyle{++}}C_{\scriptscriptstyle{++}}-\Gam_{\scriptscriptstyle{--}}C_{\scriptscriptstyle{--}}\r)=0\\
&&\Gam_{\scriptscriptstyle{-+}}\l(1-\Gam_{\scriptscriptstyle{++}}C_{\scriptscriptstyle{++}}-\Gam_{\scriptscriptstyle{--}}C_{\scriptscriptstyle{--}}\r)=0\\
&&\Gam_{\scriptscriptstyle{--}}-\Gam_{\scriptscriptstyle{--}}^{2}C_{\scriptscriptstyle{--}}-\Gam_{\scriptscriptstyle{-+}}\Gam_{\scriptscriptstyle{+-}}C_{\scriptscriptstyle{++}} =0.
\enn

First, we consider the case with the absence of inter-valley scattering, given by $\Gam_{\scriptscriptstyle{+-}}=\Gam_{\scriptscriptstyle{-+}}=0$. Then, we obtain
\bnn
\Gam_{\scriptscriptstyle{++}}-\Gam_{\scriptscriptstyle{++}}^{2}C_{\scriptscriptstyle{++}}=0 ~ \& ~ \Gam_{\scriptscriptstyle{--}}-\Gam_{\scriptscriptstyle{--}}^{2} C_{\scriptscriptstyle{++}}=0.
\enn
$(\Gam_{\scriptscriptstyle{++}},\Gam_{\scriptscriptstyle{--}})=\l\{(0,0),~(0,1/C_{\scriptscriptstyle{--}}),~(1/C_{\scriptscriptstyle{++}},0)\r\}$ are unstable fixed points, and $(\Gam_{\scriptscriptstyle{++}},\Gam_{\scriptscriptstyle{--}})=(1/C_{\scriptscriptstyle{++}},1/C_{\scriptscriptstyle{--}})$ is the only stable fixed point. This stable fixed point is described by the first self-consistent Born approximation without inter-valley scattering, where spin chirality is well defined.

Next, we consider the presence of inter-valley scattering, given by $\Gam_{\scriptscriptstyle{+-}}=\Gam_{\scriptscriptstyle{-+}}$ and $\Gam_{\scriptscriptstyle{+-}}\neq0$. Then, we obtain 
\bnn
&&\Gam_{\scriptscriptstyle{++}}C_{\scriptscriptstyle{++}}+\Gam_{\scriptscriptstyle{--}}C_{\scriptscriptstyle{--}}=1\\
&&\Gam_{\scriptscriptstyle{++}}-\Gam_{\scriptscriptstyle{++}}^{2}C_{\scriptscriptstyle{++}}-\Gam_{\scriptscriptstyle{+-}}^{2}C_{\scriptscriptstyle{--}}=0\\
&&\Gam_{\scriptscriptstyle{--}}-\Gam_{\scriptscriptstyle{--}}^{2}C_{\scriptscriptstyle{--}}-\Gam_{\scriptscriptstyle{+-}}^{2}C_{\scriptscriptstyle{++}}=0.
\enn
Solving these equations, we find two fixed points:
%
%
\bnn
(\Gam_{\scriptscriptstyle{++}},\Gam_{\scriptscriptstyle{--}})&=&\l(\f{1+\sqrt{1-4C_{\scriptscriptstyle{++}}C_{\scriptscriptstyle{--}} \Gam_{\scriptscriptstyle{+-}}^{2}}}{2C_{\scriptscriptstyle{++}}},\f{1-\sqrt{1-4C_{\scriptscriptstyle{--}}C_{\scriptscriptstyle{++}}\Gam_{\scriptscriptstyle{+-}}^{2}}}{2C_{\scriptscriptstyle{--}}}\r)\\
(\Gam_{\scriptscriptstyle{++}},\Gam_{\scriptscriptstyle{--}})&=&\l(\f{1-\sqrt{1-4C_{\scriptscriptstyle{++}}C_{\scriptscriptstyle{--}} \Gam_{\scriptscriptstyle{+-}}^{2}}}{2C_{\scriptscriptstyle{++}}},\f{1+\sqrt{1-4C_{\scriptscriptstyle{--}}C_{\scriptscriptstyle{++}}\Gam_{\scriptscriptstyle{+-}}^{2}}}{2C_{\scriptscriptstyle{--}}}\r) .
\enn 
When $\Gam_{\scriptscriptstyle{+-}}$ satisfies $1-4C_{\scriptscriptstyle{++}}C_{\scriptscriptstyle{--}} \Gam_{\scriptscriptstyle{+-}}^{2} \approx 0$, we find that these two fixed points emerge into $(\Gam_{\scriptscriptstyle{++}},\Gam_{\scriptscriptstyle{--}}) \approx (1/C_{\scriptscriptstyle{++}},1/C_{\scriptscriptstyle{--}})$. This fixed point is described by the second solution of the Born approximation in the presence of inter-valley scattering, in which the spin chirality is smeared out. This may be regarded as an intermediate solution with spin chirality before the ``topological phase transition" toward normal diffusive Fermi liquids without spin chirality appears. Table. \ref{table1} summarizes our results, where ``WAL" and ``WL" represent weak antilocalization and weak localization, respectively. 

%
%

\begin{table}[ht]
\centering
\caption{Two ground states of self-consistent Born approximation and two fixed points of the renormalization group analysis.}
\renewcommand{\arraystretch}{1.4}
\begin{tabular}{>{\centering}m{2.1in} >{\centering}m{2.1in} >{\centering\arraybackslash}m{1.6in}}
\hline
& Diffusive Helical Fermi Liquid & Diffusive Fermi Liquid \\
\hline
Fixed point    &    \shortstack[c]{~\\$\Gam_{\scriptscriptstyle{++}}\neq0$,~$\Gam_{\scriptscriptstyle{--}}\neq0$,\\and $\Gam_{\scriptscriptstyle{+-}}=0$}    &    \shortstack[c]{$\Gam_{\scriptscriptstyle{++}}\neq0$,~$\Gam_{\scriptscriptstyle{--}}\neq0$,\\and $\Gam_{\scriptscriptstyle{+-}}\neq0$} \\
\shortstack[c]{~\\Ground State\\~~(Self-consistent Born analysis)~~}   &   \shortstack[c]{~\\$Q_{\scriptscriptstyle{++}}\neq0$,~$Q_{\scriptscriptstyle{--}}\neq0$,\\and $Q_{\scriptscriptstyle{+-}}=0$}   &   \shortstack[c]{~\\$Q_{\scriptscriptstyle{++}}\neq0$,~$Q_{\scriptscriptstyle{--}}\neq0$,\\and $Q_{\scriptscriptstyle{+-}}\neq0$}\\
Transport property   &    WAL   &   \shortstack[c]{~\\WAL$\rightarrow$WL\\(Crossover)}\\
\hline
\end{tabular}
\label{table1}
\end{table}


\section{Discussion}
Considering that the only relevant energy scales are the cyclotron energy $\hbar \omega$ and the Fermi energy $E_F$ in the IFS, it is natural to introduce a single parameter $b =  \hbar \omega/E_F = (\hbar e/m_{WF}E_F)B$ for the Hall resistivity contribution from the Weyl fermions, anticipating the scaling behavior for $\Delta \rho_H$ \cite{HJKim11}, where $m_{WF}$ is an effective mass of the Weyl fermion and $\Delta \rho_H$ is the Hall resistivity component deviating from the linearity. Indeed, we found a scaling property in $\Delta \rho_H$, presented in Fig. 9(a), where the y-axis should be also scaled as the magnitude of $\Delta \rho_H$ is inversely proportional to the carrier density. This scaling analysis enables us to estimate $m_{WF}$, whose values for all six samples are plotted in Fig. 9(b) as a function of the corresponding  $E_F$.  $m_{WF}$ is in the order of 
$10^{-6} - 10^{-4} m_0$, where $m_0$ is the mass of an electron and they exhibit a singular behavior with a minimum at the Weyl point.

It is straightforward to find  $m_{WF} = m - \frac{\alpha_R}{\sqrt{\alpha_R^2+\frac{2 \hbar^2}{m}E_F}}m$ from the Rashba Hamiltonian with degenerate parabolic bands, where $m$ is the bare band mass given by the curvature of the parabolic band and the Rashba coupling constant $\alpha_R$ determines the energy of the Weyl point from the bottom of the conduction band, given by $E_{WF} =\frac{1}{2}\frac{m}{\hbar^2}\alpha_R^2$. Considering an overall shift for the Fermi energy and taking the limit of $\alpha_R^2 >> \frac{2 \hbar^2}{m}E_F$, we obtain $m_{WF} \approx \frac{\hbar^2}{\alpha_R^2} \mid E_{F}-E_{WF} \mid$. This equation describes zero mass at the Weyl point quite well. However, compared to the experimental result, the mass increases rather steeply as $E_F$  deviates from the Weyl point. In the opposite limit of the strictly linear dispersion, the mass is zero even away from the Weyl point. According to the density functional theory (DFT) \cite{Bahramy11}, the dispersion near the Weyl point is neither quadratic nor linear. Therefore, to parameterize the degree of deviation from the linear dispersion, we introduce a phenomenological equation,
$\frac{m_{WF}}{m} \approx \frac{1}{2} | 1-(\frac{v_{linear}}{v_{real}} )^2 |\frac{\mid E_F-E_{WF} \mid}{E_{WF}}$. 
Assuming $\frac{v_{linear}}{v_{real}} \approx 1+\varepsilon$, we obtain $\varepsilon \sim 2.2 \times 10^{-2}$. This result implies that all higher order terms of the curvature in the dispersion is only few \% and thus, the real dispersion in BiTeI near the Weyl point is considerably linear. Though the dispersion is mainly determined by periodic ionic potentials, we do not exclude any contribution to the linear dispersion resulting from electron interaction.

The universal scaling of the Hall resistivity discussed above is quite consistent with the extreme disparity of the mobility and divergent IFS mobility. Representing the Hall resistivity $\Delta \rho_H(B) =\frac{1}{nec} \frac{B}{1+\mu^2B^2}$  as
 $\Delta \rho_H(b)=\frac{1}{nec}\frac{\frac{m_{WF}E_F}{\hbar e}b}{1+\mu^2(\frac{m_{WF}E_F}{\hbar e})^2b^2}$  with the dimensionless magnetic field   $b$ discussed before, we obtain the scaling expression of $\frac{\Delta \rho_H(b)}{\Delta \rho_H(b=1)}=\frac{(1+\mu_{sc}^2)b}{1+\mu_{sc}^2b^2}$, 
 where $\mu_{sc}=\frac{m_{WF}E_F}{\hbar e}\mu$  is a scaled mobility. This scaled mobility, being a universal constant does not depend on the Fermi energy. Introducing the empirical formula introduced above for Weyl-fermion mass into the mobility, we find the following expressions of IFS and OFS mobility, given by $\mu_{IFS}(E_F)=2\frac{\hbar e \mu_{sc}}{m|1-(\frac{v_{linear}}{v_{real}})^2|} \frac{E_F}{|E_F||E_F-E_{WF}|}$  
 and $\mu_{OFS}(E_F)=\frac{\hbar e\mu_{sc}}{2m|E_F|}$, respectively. As $E_F$  is inversely proportional to $m$  for the charge carrier on OFS,
 $\mu_{OFS}$   is constant and $\mu_{IFS}$  follows $\mu_{IFS}(E_F) \propto \frac{1}{|E_F-E_{WF}|}$ 
 with the ratio of $\frac{\mu_{IFS}(E_F)}{\mu_{OFS}(E_F)}=\frac{m_{OFS}}{m_{WF}}=\frac{4}{|1-(\frac{v_{linear}}{v_{real}})^2|}\frac{E_{WF}}{|E_F-E_{WF}|} \approx 10^3 \sim 10^4$. Indeed, this scaling argument is consistent with the ``divergent'' $\mu_{IFS}$  at the Weyl point shown in the experimental result [Fig. 1(b)]. Note that while the mass ratio between IFS and OFS in this argument is mostly determined by the empirical factor
 $\varepsilon \approx1-\frac{v_{linear}}{v_{real}}$  introduced above, the ``divergent'' behavior of $\mu_{IFS}$  at the Weyl point is given by 
 $\frac{1}{|E_F-E_{WF}|}$  and in fact, this term is inherent in the Rashba model.

In order to understand the origin of the divergent IFS mobility and the extreme disparity between IFS and OFS mobility near the Weyl point, 
we have performed the self-consistent Born analysis for the Rashba Hamiltonian, which is a mean-field theory in the presence of disorder with no consideration of electron correlation. Here we summarize main results of the perturbative renormalization group analysis to understand how a fixed-point phase is determined. In general, disorder strength increases at the short-distance scale in three dimensions because huge number of electrons on the Fermi surfaces are affected by disorder potentials. On the other hands, it decreases at the long-distance scale because disorder potentials are effectively screened. As a result, balancing is achieved and it gives rise to a finite-disorder fixed point, which is known as a diffusive Fermi liquid. In the present problem, we found two types of fixed points: One contains the effects of the intra-valley forward scattering only and the other considers both intra- and inter-valley scattering. We have performed the self-consistent Born approximation and found a fixed-point solution for the electron Green’s function in both cases. Then, we have calculated transport coefficients, evaluating current-current correlation functions with this mean-field-theory propagator. 

Fig. 8 shows that the self-consistent Born analysis describes our experimental data quantitatively, where lines and discrete points represent theoretical curves and experimental results, respectively. It is natural to expect that the presence of inter-valley scattering reduces the mobility. However, effects of the inter-valley scattering are not relevant in describing the experimental data in the present case
 because it suppresses spin chirality and the weak anti-localization. This fixed point is distinguished from a conventional diffusive Fermi liquid because of definite chirality  and we name it a diffusive helical Fermi liquid. Thus, our BiTeI single crystals are weakly disordered with negligible inter-valley scattering whose ground state is considered to be a diffusive helical Fermi liquid. We would like to emphasize that only one fitting parameter, related with the variance of disorder potential at the fixed point is used in this comparison, whereas all other parameters are determined by the experiment.

It is straightforward to understand the divergent IFS mobility within the framework of the self-consistent Born approximation. As the IFS density of states vanishes, approaching the Weyl point, the scattering rate also becomes zero at the Weyl point. However, it is difficult to explain the experimentally confirmed scaling of Hall resistivity within the same framework. In fact, we find that scaled mobility 
$\mu_{sc} = \frac{m_{WF}E_F}{\hbar e} \mu$  is not independent of the Fermi energy $E_F$  when the mobility $\mu$  evaluated from the self-consistent mean-field analysis is used. The independence of $\mu_{sc}$ on $E_F$ is achieved only when the divergence of the IFS mobility is exactly cancelled by the mass reduction near the Weyl point. In the scaling argument, we obtained $m_{WF} \propto |E_F - E_{WF}|/E_F$. On the other hand, Born mean-field theory gives $\mu_{IFS}(E_F) \propto \frac{1}{|E_F-E_{WF}|^{\kappa}}$  with  $\kappa$ larger than 1. Thus, the $\mu_{sc}$ 
 is not independent of $E_F$  in the self-consistent mean-field analysis.
 
 One way to reconcile this inconsistency is to take into account the role of effective interactions between electrons near the Weyl point. As the IFS density of states vanishes at the Weyl point, effective interactions can be enhanced due to weaker screening effect. In fact, this is what happens in graphene. Correlation effects indeed reshape the linear band dispersion of graphene \cite{Elias11,Siegel11,Kotov12,Chae12}. Possible interplay among inversion symmetry breaking (spin chirality), disorders, and effective interactions may lead to a novel interacting diffusive fixed point, which allows the universal scaling in the Hall resistivity.
 
When disorders become stronger, it is possible that a topological structure (geometric phase) in the ground-state wave-function changes. Previously, we considered two types of fixed points, corresponding to the absence and presence of inter-valley scattering, respectively.  The former solution would be justified when effects of disorder scattering are not strong, called a diffusive helical Fermi-liquid state. On the other hand, the second solution is expected to be realized when disorder scattering becomes more relevant than the first, identified with a diffusive Fermi-liquid state. A crucial difference is spin chirality. Intra-valley scattering preserves the spin chirality while inter-valley scattering destroys it. As a result, we predict that the weak anti-localization turns into the weak localization in the presence of strong inter-valley scattering, increasing disorder strength. This crossover behavior from the diffusive helical Fermi liquid to the conventional diffusive Fermi liquid may be regarded as a weak version of a topological phase transition driven by disorder although BiTeI is topologically trivial.

\section{Conclusion}
In conclusion, we uncovered that the interplay between disorder and inversion symmetry breaking is responsible for (1) divergent mobility in the inner chiral Fermi surface (FS), (2) extreme disparity of the mobility values between the inner and outer chiral FS, and (3) universal scaling in the Hall resistivity. Based on the self-consistent Born approximation, we could consistently explain the observation (1) and (2), quantitatively reproducing mobility values of the inner and outer FS as a function of the Fermi energy. However, the universal scaling of the Hall resistivity cannot be accounted for within this mean-field theory, which indicates the existence of mass renormalization of the inner Fermi-surface near the Weyl point, possibly originated from electron correlation due to weaker screening near the Weyl point. \\

\acknowledgements
This study was supported by Basic Science Research Program through the National Research Foundation of Korea (NRF) funded by the Ministry of Education, Science, and Technology (No. 2014R1A1A1002263). KS was also supported by the Ministry of Education, Science, and Technology (No. 2012R1A1B3000550 and No. 2011-0030785) of the National Research Foundation of Korea (NRF) and by TJ Park Science Fellowship of the POSCO TJ Park Foundation. MS was also supported by YO-COE Foundation from Yamagata University. MS wishes to express his thanks to Prof. T. Iwata for his support. 

$^{\ast}$ Both authors equally contribute to the present paper \\
$^{\dagger}$ Corresponding author; hjkim76@daegu.ac.kr

\bf
\caption{(a) Schematic picture of the spin-split band of BiTeI at the A point in the reciprocal space. (b) The mobility of the electrons on the OFS, $\mu$  and that of the Weyl fermions in the IFS, $\mu_{WF}$ as a function of the corresponding Fermi energy $E_F$. 
The red (black) solid and open squares (circles) are the values of $\mu_{WF}$ ($\mu$) obtained from the magnetoconductivity and the Hall resistivity, respectively. The red (black) line is a theoretical curve for $\mu_{WF}$ and $\mu$ given by $1/|E_F-E_{WF}|$ and a constant. The $E_F$-dependence of $\mu_{WF}$ and $\mu$ is explained in the Sec. IV. The dashed (red) line is the approximate position of the Weyl point. }
\label{fig1}
\ef

\bf
\caption{Temperature dependence of the resistivity $\rho$($T$) for the sample $\#1 - \#6$. }
\label{fig2}
\ef

\bf
\caption{(a) The deviation of the Hall resistivity from the linearity $\Delta\rho_H$, which clearly 
shows an anomalous feature of Hall resistivity $\rho_H$($B$) at low magnetic fields. (b) Magnetoresistance 
of the sample $\#1 - \#6$ for -4 T $< B <$ 4 T.  }
\label{fig3}
\ef

\bf
\caption{ The fitting of $\Delta\rho_H$($B$) for the sample $\#1$ and $\#6$ based on Eq. (1) in the text.  }
\label{fig4}
\ef

\bf
\caption{ (a) Decomposition of conductivity contributions for the sample $\#1$. The solid circle is the magnetoconductivity $\Delta\sigma$, where the $B$-independent constant is subtracted. The (blue) dashed line represents the conductivity contributions from electrons in the OFS. The open circle is $\Delta\sigma-\Delta\sigma_{OSF}$. The (red) solid line is the theoretical fitting to $\Delta\sigma-\Delta\sigma_{OSF}$,
 based on the orbital contribution $\Delta\sigma_{WF,o}$ and the weak antilocalization  $\Delta\sigma_{WAL}$.
 (b)The scaled contributions of the weak antilocalization $\Delta\sigma_{WAL}$  with $\sqrt{b}$  dependence, where $b$ is the scaled magnetic field
 given by $b=\hbar\omega/E_F$. }
\label{fig5}
\ef

\bf
\caption{Evolution of Fermi surfaces in BiTeI as a function of the Fermi energy $E_F$. 
The middle figure shows the projection of Fermi surface on the $k_x-k_z$ plane when the $E_F$ crosses right at the Weyl point. 
The left (right) figure shows the cases when the $E_F$ is above (below) the Weyl point. 
A three dimensional image can be constructed by rotating this projection along the $z-$direction.}
\label{FS_3d}
\ef

\bf
\caption{Comparison between the results from a theory based on the self-consistent Born approximation without inter-valley scattering
 and experimental data. Two fitting parameters have been used as $b=8.85$ $[(eV)^{-1}]$ and $A=0.984$ $[m^{2}/Vs]$. $E_{F}$ is shifted by $1/b$ for the formula to be matched with the experiment, where the conduction band minimum is at 0 eV. The red line represents the diffusion constant ($D_{+}$) for the inner Fermi surface 
and the green line describes  ($D_{-}$) for the outer Fermi surface. As the Fermi energy approaches the Weyl point, where the vertical line is located ($E_{F}$ = 113 meV), $D_{+}$ shows a divergent behavior but $D_{-}$ remains almost unchanged. }
\label{diffusion_constant_3d}
\ef

\bf
\caption{Comparison between the results from a theory based on self-consistent Born approximation with inter-valley scattering and experimental data. Two fitting parameters have been used as $b=8.85$ $[(eV)^{-1}]$ and $A=0.984$ $[m^{2}/Vs]$, the same as before. Dashed lines of the result without inter-valley scattering are drawn for comparison. The off-diagonal component lowers mobilities slightly for both fermions.}
\label{diffusion_constant_3d_with_offQ}
\ef

\bf
\caption{ (a) The scaling behavior of $\Delta \rho_H$. The (red) solid line is a guide to eyes. (b) the effective mass $m_{WF}$ of the Weyl fermions, deduced from the scaling in (a) at the corresponding $E_F$.}
\label{fig9}
\ef



\begin{thebibliography}{9}
\bibitem{Mathematica} We find such elliptic integrals using Mathematica.
\end{thebibliography}
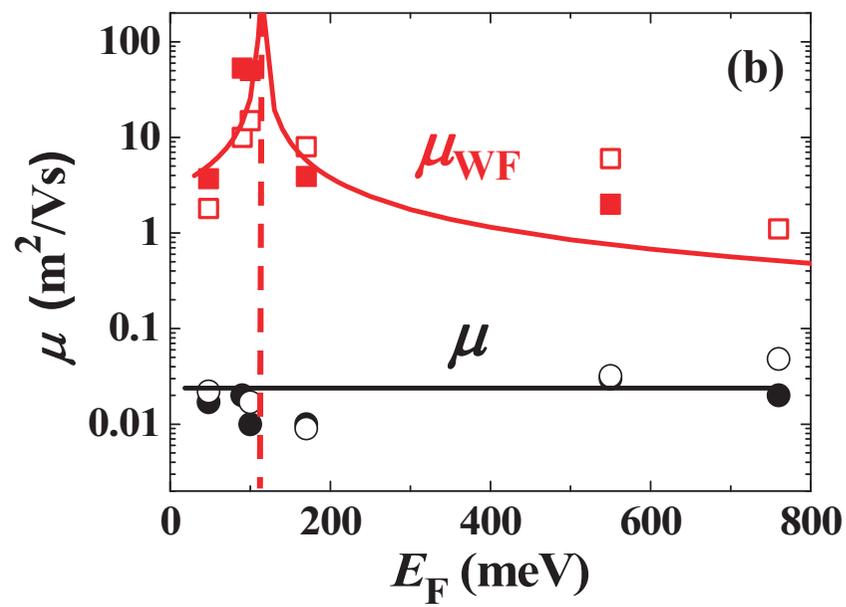
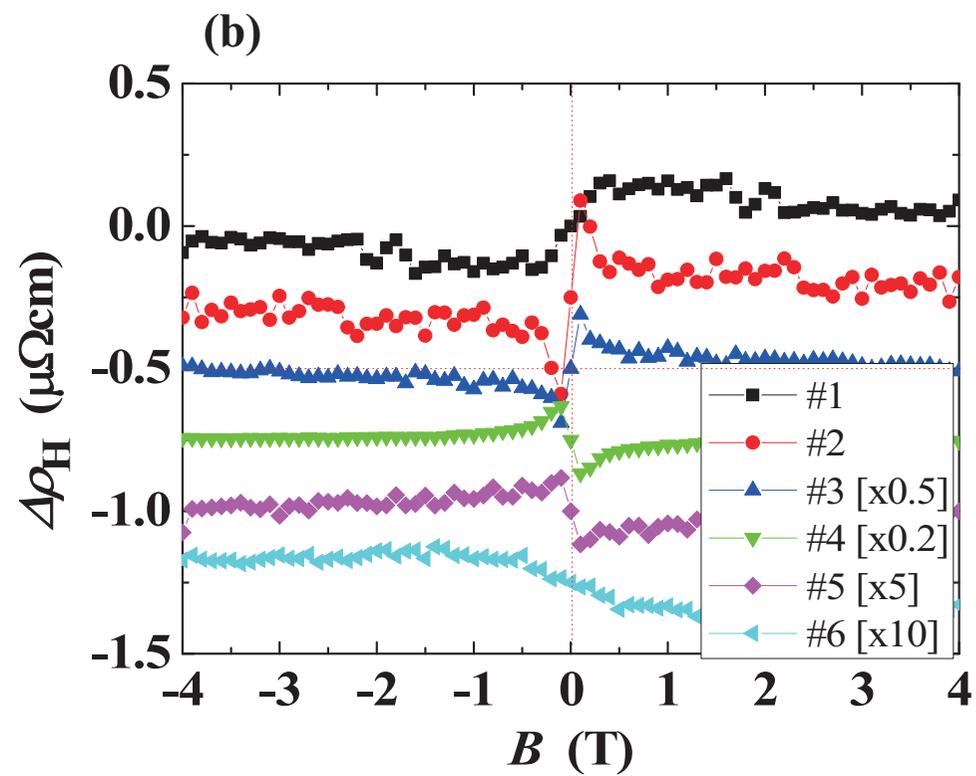
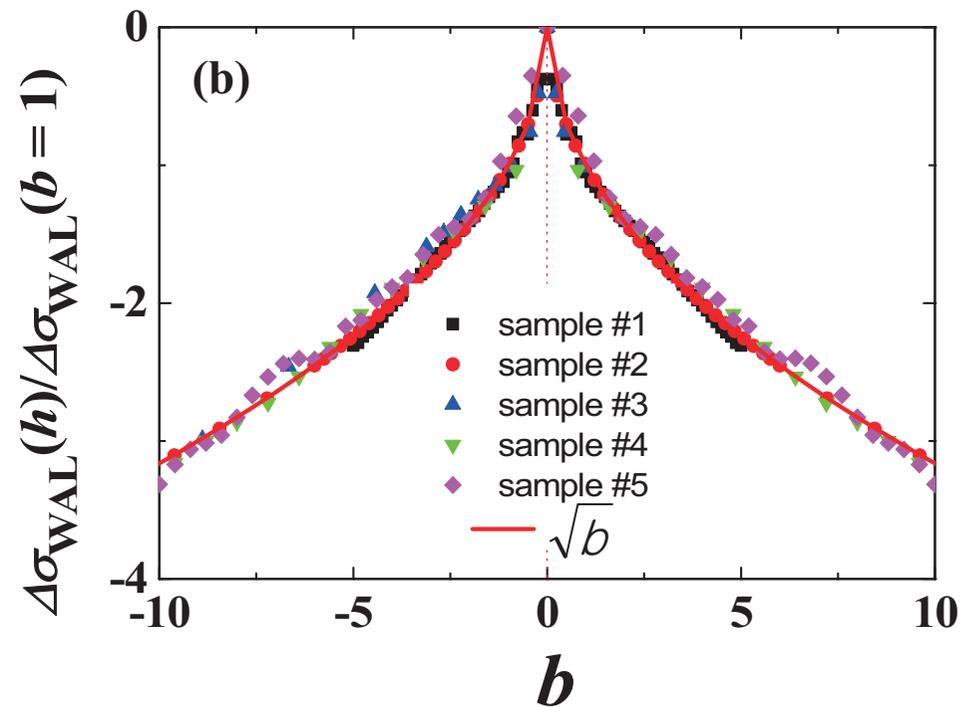
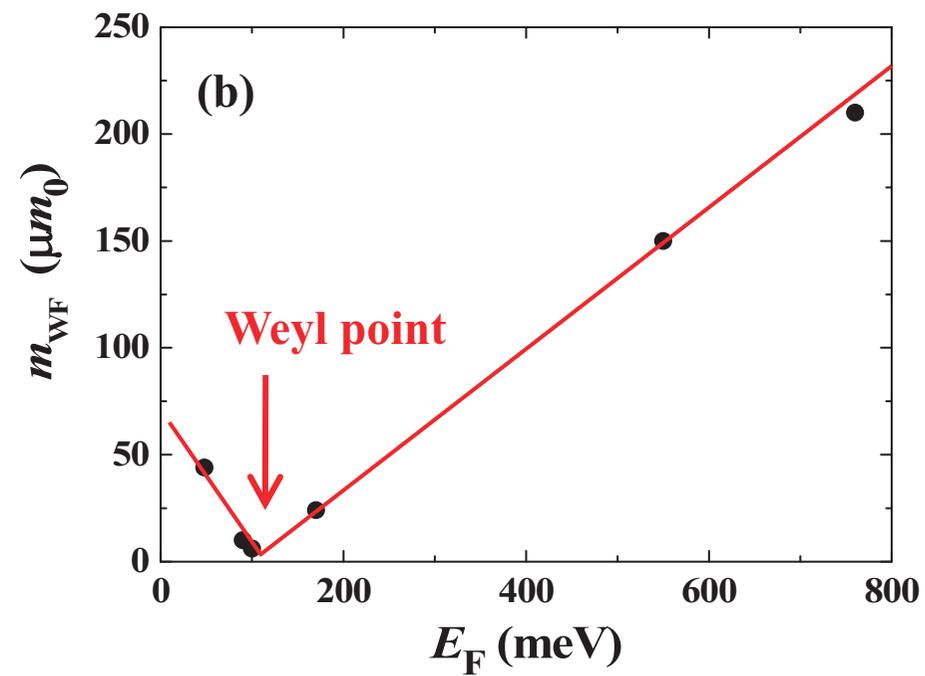
\end{document}